\documentclass[aps, twocolumn, superscriptaddress, longbibliography]{revtex4-1}  % for review and submission
\usepackage{graphicx}  % needed for figures
\usepackage{dcolumn}   % needed for some tables
\usepackage{bm}        % for math
\usepackage{amssymb}  
\usepackage{hyperref} % for math
\usepackage{xr}
\usepackage{url}
\usepackage{color}
\usepackage{cancel}
\usepackage{amsmath}
\usepackage{epsfig}
\usepackage{epstopdf}
\newcommand{\Kij}{K_{ij}^{\alpha}}
\newcommand{\Kkm}{K_{km}^{\beta}}
\newcommand{\ai}{a_{i}^{\alpha}}
\newcommand{\ak}{a_{k}^{\beta}}
\newcommand{\pr}{\partial}
\newcommand{\s}{\displaystyle \sum}
\makeatletter 

\begin{document}

\title{Spin models inferred from patient data faithfully describe HIV fitness landscapes and enable rational vaccine design}

\author{Karthik Shekhar}
\affiliation{Department of Chemical Engineering, MIT, Cambridge, MA 02139} 
\affiliation{Ragon Institute of MGH, MIT and Harvard, Boston, MA 02129}

\author{Claire F. Ruberman}
\affiliation{Department of Mathematics, Pomona College, Claremont, CA 91711}

\author{Andrew L. Ferguson}
\affiliation{Department of Materials Science and Engineering, University of Illinois at Urbana-Champaign, Urbana, IL 61801}

\author{John P. Barton}
\affiliation{Department of Chemical Engineering, MIT, Cambridge, MA 02139} 
\affiliation{Ragon Institute of MGH, MIT and Harvard, Boston, MA 02129}

\author{Mehran Kardar}
\thanks{Corresponding author}
\email{kardar@mit.edu}
\affiliation{Department of Physics, MIT, Cambridge, MA 02139}

\author{Arup K. Chakraborty}
\thanks{Corresponding author}
\email{arupc@mit.edu}
\affiliation{Department of Chemical Engineering, MIT, Cambridge, MA 02139} 
\affiliation{Ragon Institute of MGH, MIT and Harvard, Boston, MA 02129}
\affiliation{Department of Physics, MIT, Cambridge, MA 02139} 
\affiliation{Department of Chemistry, MIT, Cambridge, MA 02139} 
\affiliation{Department of Biological Engineering, MIT, Cambridge, MA 02139} 
\affiliation{Institute for Medical Engineering and Science, MIT, Cambridge, MA 02139}

\pagebreak
\begin{abstract}
Mutational escape from vaccine induced immune responses has thwarted the development of a successful vaccine against AIDS, whose causative agent is HIV, a highly mutable virus.  Knowing the virus' fitness as a function of its proteomic sequence can enable rational design of potent vaccines, as this information can focus vaccine induced immune responses to target mutational vulnerabilities of the virus. Spin models have been proposed as a means to infer intrinsic fitness landscapes of HIV proteins from patient-derived viral protein sequences. These sequences are the product of non-equilibrium viral evolution driven by patient-specific immune responses, and are subject to phylogenetic constraints.  How can such sequence data allow inference of intrinsic fitness landscapes? We combined computer simulations and variational theory \'{a} la Feynman to show that, in most circumstances, spin models inferred from patient-derived viral sequences reflect the correct rank order of the fitness of mutant viral strains.  Our findings are relevant for diverse viruses.
\end{abstract}

\maketitle

\section{Introduction}

The staggering sequence diversity of HIV \cite{gaschen2002diversity} and its ability to evade most natural or vaccine-induced immune responses by mutational escape \cite{goulder2004escape} have precluded the development of a successful vaccine against this global epidemic \cite{walker2008vaccine}. It has been proposed that a vaccine-induced immune response should target regions in the viral proteome where escape mutations are most likely to damage replicative fitness.  Single residues that appear highly conserved in proteins derived from virus samples extracted from diverse patients have been suggested as vaccine targets \cite{letourneau2007vaccine}, but the fitness cost \cite{schneidewind2007escape} of making mutations at such sites can be restored by additional compensatory mutations \cite{troyer2009variable}.  Groups of sites in HIV proteins that are collectively constrained such that multiple simultaneous mutations within such groups impose a high fitness penalty have been identified, and shown to be targeted by patients whose immune systems naturally control HIV \cite{dahirel2011coordinate}. But these models cannot identify which specific sites in these collectively co-evolving groups should be targeted to maximally compromise viral fitness, and how mutational escape pathways that exist even within these regions may be blocked by additional immune responses. Answering these questions requires knowledge of the complex, multidimensional structure of HIV's fitness landscape \cite{kouyos2012landscape, ferguson2012hivtrap} - a measure of the virus' replicative capacity as a function of the amino acid sequence of its constituent proteins. Knowledge of the fitness landscape can guide systematic identification of the mutational vulnerabilities of viruses (not just HIV), and the rational design of vaccines that can target these weaknesses.

Using inference principles rooted in entropy maximization \cite{jaynes1957infotheory, tkacik2009spin}, Ferguson et al. \cite{ferguson2012hivtrap} recently employed publicly available multiple sequence alignments (MSA) \cite{hivdb} of four HIV proteins to obtain the prevalence of HIV strains bearing multiple mutations in these proteins. In this approach, a viral protein of $N$ sites is described by a coarse-grained binary code $\vec{s} = \{0,1\}^N$, wherein the ``wild-type" (most frequent) amino acid at site $i$ is denoted by $s_i=0$ and any mutant is denoted by $s_i=1$ (irrespective of its identity).  Given the protein MSA, the maximum entropy framework seeks a minimally biased probability distribution $P[\vec{s}\,]$ over the space of all possible mutant strains $\{\vec{s}\,\}$ that reproduces the marginal one-site and two-site mutational probabilities $\langle s_i \rangle_{MSA}$ and $\langle s_i s_j \rangle_{MSA}$ measured from sequence data. The resulting inference leads to a model where the probability of a particular strain $\vec{s}$, the \textit{prevalence landscape}, is described by a Boltzmann distribution, $P[\vec{s}\,] \propto e^{-H_0[\vec{s}\,]}$, where the Hamiltonian takes the form of an infinite-range Ising spin glass \cite{binder1986spin},

\begin{equation}
H_0[\vec{s}\,] = \sum_{i<j = 1}^N J_{ij} s_i s_j +  \sum_{i=1}^N h_i s_i.
\label{maxent}
\end{equation}
This can be generalized to account for the identities of mutant amino acids using Potts models rather than an Ising model \cite{ferguson2012hivtrap}.

Only under certain restricted circumstances~\cite{sella2005stat} can it be rigorously proven that this prevalence landscape is the replicative fitness landscape, with the ``energy" $H_0[\vec{s}\,]$ of a strain $\vec{s}$ being inversely correlated with its replicative capacity. But, these conditions are likely inapplicable for a complex problem like HIV evolution in human patients, and the connection between fitness and prevalence is not obvious. Ferguson et al.~\cite{ferguson2012hivtrap} report a statistically significant negative correlation between values of $H_0[\vec{s}\,]$ predicted by the inferred model and \textit{in vitro} fitness measurements of several engineered strains (including both published and \textit{de novo} experiments). Their predictions also tested positively against clinical data.  These results provide evidence that the prevalence landscape of the virus described by $H_0[\vec{s}\,]$ is a good proxy for the intrinsic fitness landscape.

This pleasing result is surprising. However, the quantities used to parametrize the prevalence landscape ($\langle s_i \rangle_{MSA}$ and $\langle s_i s_j \rangle_{MSA}$) were obtained from ``consensus" (most common) protein strains in genetically diverse patients.  The sequence evolution of HIV within a particular host is a non-equilibrium process driven by a genetically determined pattern of \textit{immune  pressure} acting on the viral proteome~\cite{brumme2009pathway}, which determines the effectively fittest viral strains in each patient. Thus, the HIV sequences used to infer the prevalence landscape were not sampled from an equilibrium ensemble of sequences distributed according to their ``intrinsic" fitness. In contrast, the \textit{in vitro} measurements are not subject to human immune responses, and therefore assay the intrinsic fitness of the virus. The robust correlation between model predictions and \textit{in vitro} fitness measurements observed by Ferguson et al.~\cite{ferguson2012hivtrap}, and their ability to describe clinical data in humans with diverse genotypes when the immune responses were known, therefore pose an important question: How does a prevalence landscape inferred from the statistics of mutations in a non-equilibrium ensemble of sequences evolving under diverse adaptive immune responses faithfully reflect the intrinsic fitness of mutant viral strains?      

Here we combine computer simulations and analytical theory to address this question. We find that the presence of genetically diverse immune responses imposed by patients across the population is necessary for comprehensive sampling of the sequence space of viral proteins. We then show that the prevalence landscape inferred from mutational correlations observed in the sequence databases correctly reflects the  rank order of the intrinsic replicative fitness of mutant viral strains in most circumstances. We provide mechanistic insights into why this is so, and circumstances wherein this may not be true. 

\section{Simulations}

\begin{figure*}[htp!]
\centering
\includegraphics[scale=0.6]{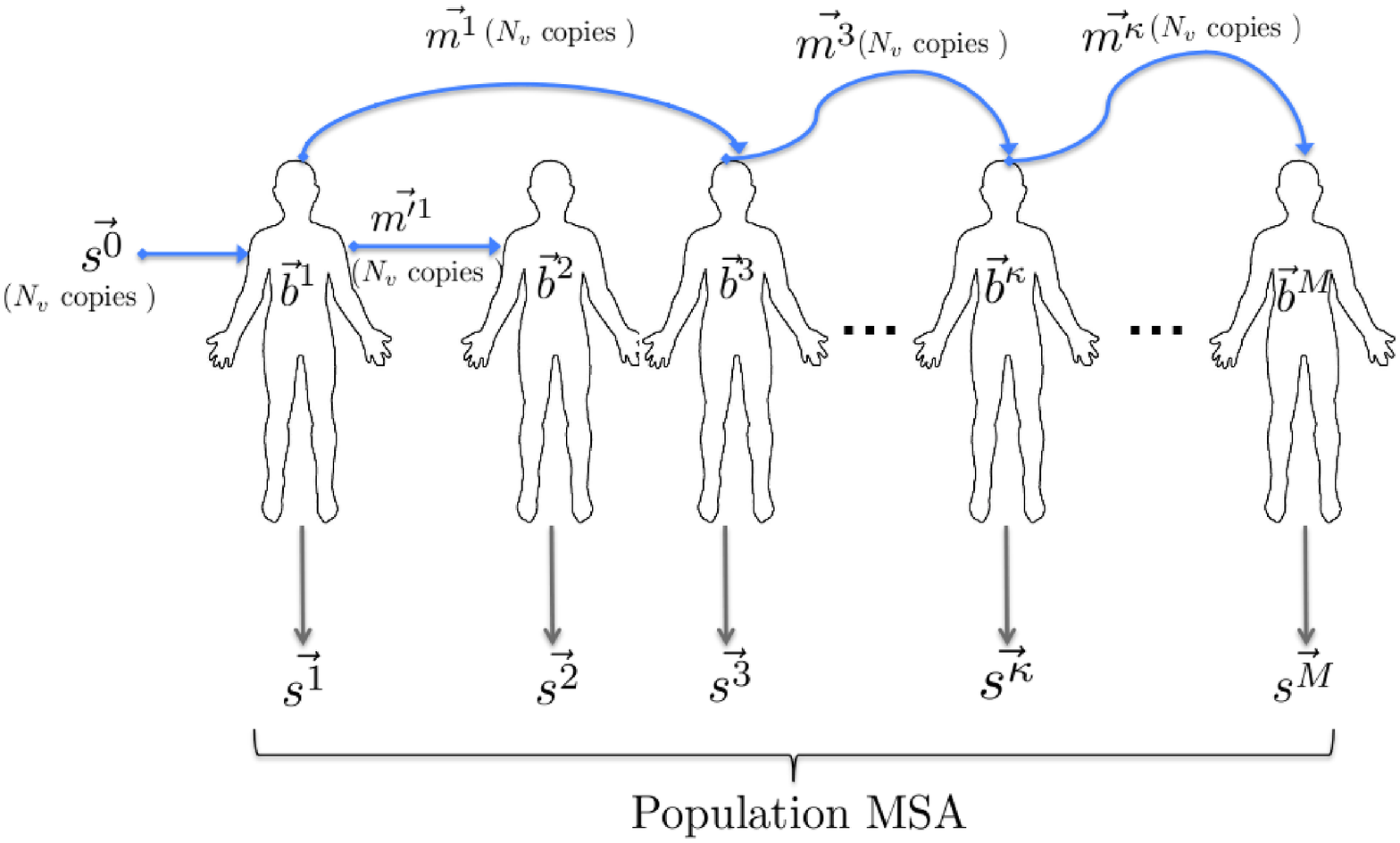}
\caption{\textbf{Graphical description of simulation model}\\
We consider an expanding network of infected hosts. The first host in the network is infected with $N_v$ copies of $\vec{s^{0}} = \vec{0}$, corresponding to the ``wild-type" (WT) strain. Each new host added to the network is infected with $N_v$ copies of a single viral strain derived from the quasispecies within an existing host chosen at random (based on evidence that most infections are initiated by a single virus strain).  $\vec{b^{\kappa}}$ is a ``field" that acts on the proteomic sites within host $\kappa$ and represents the genetically determined immune response within that host. The ``consensus" viral strain is extracted from every productively infected host $i\,\in \,1,\,2,\,\ldots,\,M$, and added to the \textit{in silico} population ensemble, mimicking the way these sequences were sampled from a real population. For example, in this figure, infection in host $\kappa$ is seeded with $N_v$ copies of strain $\vec{m^{3}}$, which is randomly chosen from the quasispecies within host 3. At a randomly chosen generation of viral quasispecies evolution within host $\kappa$, the consensus strain $\vec{s^{\kappa}}$ is derived from the quasispecies and added to the population ensemble.  }
\label{fig1}
\end{figure*}

For the computer simulations, as an example, we study the 132-residue HIV matrix protein p17~\cite{turner1999struc}. We consider a growing population of infected hosts, and model the network of viral transmission between hosts as illustrated in Fig.~\ref{fig1}. In all simulations, infection in the first host is  seeded with $N_v$ copies of the wild-type (WT) strain ($\vec{s^0}=\vec{0}$\,). New hosts added to this network are infected with $N_v$ identical copies of a strain randomly selected from the quasispecies within a host chosen randomly from the existing pool of hosts.  For simplicity, the number of viral strains in a host is chosen to be a constant $N_v$, making our intra-host evolutionary model similar in spirit to Wright-Fisher models \cite{rouzine1999linkage}. $N_v$ serves as an ``effective population size" as in conventional population genetics, and not the actual number of strains in a host, which is usually much larger than the range of $N_v$ ($2 \times 10^3 - 5 \times 10^5$) we have considered \cite{kouyos2006stochastic}.

We assume that the fitness of a particular strain $\vec{s}$ of the $N$-site protein in a given host is described by an \textit{effective} Hamiltonian $H[\vec{s}\,] = H_{int}[\vec{s}\,] + I[\vec{s}\,]$. Here, $H_{int}[\vec{s}\,] = \sum_{i<j = 1}^N J_{ij} s_i s_j + \sum_{i=1}^N h_i s_i$ constitutes intrinsic fitness that is independent of host, while $I[\vec{s}\,] = - \sum_{i=1}^N b_i s_i$ is a ``host-specific" immune pressure which applies only to some sites: the sites $s_i$ for which $b_i$ are non-zero are distinct for different hosts.  Within each host, the fields $\{ b_i \}$ are chosen in a manner consistent with known clinical information (see Appendix \ref{appA} and \emph{Supplementary Methods}). Post infection, viral quasispecies in a host evolve through a non-equilibrium  mutation-selection process in discrete generations (see Appendix \ref{appB}). In our model, the evolutionary timescale is coarse-grained such that each generation corresponds to multiple replication cycles of the viral quasispecies.  We evolve the quasispecies in a host for a random number of generations $\tau_S$, chosen uniformly between 25 and 500. Empirically, we find that our results do not change qualitatively as long $\langle \tau_S \rangle > 150$ (cf. \emph{Supplementary Methods}) which might tentatively correspond to a time scale in which the quasispecies is able sense the immune pressure and respond through adaptive mutations.  The consensus strain within each host is extracted at a randomly chosen generation, and an ensemble of such strains is recorded, which we refer to as the \emph{population ensemble}. This mimics how actual sequences in public databases were collected from patients.

For $\{J_{ij},\,h_i\}$, the parameters of $H_{int}[\vec{s}\,]$, we have used numerical values of the maximum entropy model $H_0[\vec{s}]$ of Ferguson et. al. \cite{ferguson2012hivtrap} inferred from available p17 sequences \cite{hivdb}. In other words, we assume that the intrinsic fitness landscape is correctly inferred from the database of sequences derived from patients. Recall that $H_0[\vec{s}]$ reproduces the one and two-point mutational probabilities within the real MSA. From the simulations we obtain ``virus samples'' from diverse patients. We then ask whether the mutational probabilities in this \emph{in silico} population ensemble are the same as those that describe the intrinsic fitness landscape. If they are the same, then our assumption that the intrinsic fitness landscape is described by the maximum entropy model is exactly correct. If they differ, we can evaluate what the differences are, and determine how the inferred prevalence landscape relates to an intrinsic fitness landscape. 

\section{Results}

\begin{figure*}[htp!]
\centering
\includegraphics[scale=0.4]{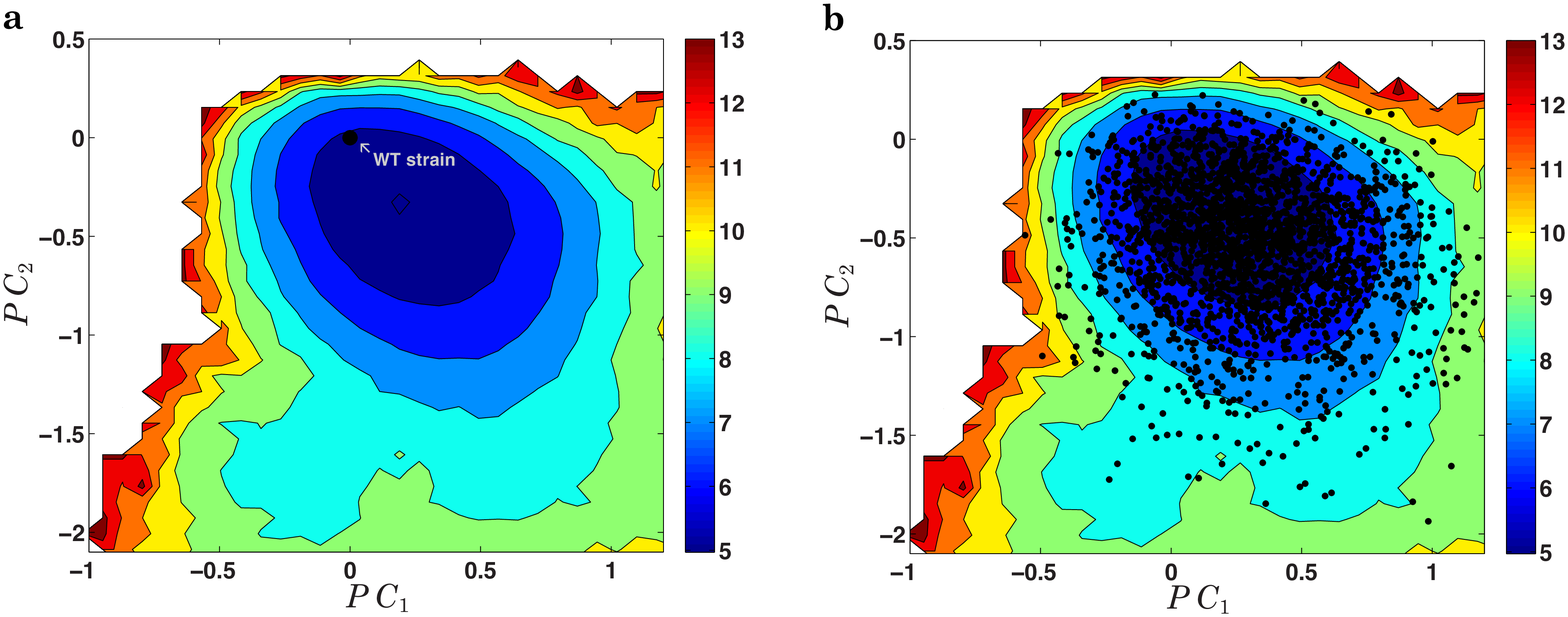}
\caption{\textbf{Immune pressure facilitates exploration of the virus in sequence space}\\ 
Viral exploration of sequence space in our simulations is depicted using a lower dimensional representation of the intrinsic fitness landscape of the protein p17, computed by applying principal component analysis (PCA) to sequences resulting from an equilibrium sampling of $H_{int}[\vec{s}\,]$ (cf. \emph{Supplementary Methods}). We only focus on the primary basin relevant to our simulations (cf. Supplementary Fig. S2). Different colors represent contours of the free energy computed as $A(x,y) = -\log P(x,y)$, where $P(x,y)$ is the normalized density of sequences at point $(x,y)$ on the $[PC_1,  PC_2]$-plane. Low values correspond to regions of high fitness. (a) In the absence of immune pressure($\{b_i^{\kappa}\} =0$), the population ensemble extracted from our simulations consists of only WT sequences which are represented by a single `$\bullet$', located at (0,0) (b)  In the presence of immune pressure, the population ensemble consists of sequences that explore different parts of the landscape. Each `$\bullet$' represents the most frequent strain in a particular host.}
\label{fig2}
\end{figure*}

To aid visualization, we computed a 2D embedding of the intrinsic fitness landscape associated with $H_{int}[\vec{s}\,]$, which charts the peaks and valleys of fitness in sequence space (see Appendix \ref{appC}).  We first simulated our model in the absence of immune pressure ($\{b_i \}=0$) for values of $\mu$ in the range $10^{-5} -10^{-2}$ /site/generation. In these simulations, the quasispecies within every host stays localized around the WT strain $\vec{s} = \vec{0}$ (Fig.~\ref{fig2}(a) and Supplementary Fig. S8).  The population ensemble in this case is entirely composed of WT strains and mutations are rarely selected at the population level. The frequencies of single and double mutations in the population ensemble, $\langle s_i \rangle_{dyn}$ and $\langle s_i s_j \rangle_{dyn}$ are close to zero (unlike in the real MSA), and reveal no information about the correlation structure of the fitness landscape.

This is because in the absence of any immune pressure, for reasonably large values of $N_v$ typical of the chronic phase of infection during which virus samples are collected, selective forces dominate genetic drift and suppress the fixation of mutations that are deleterious to intrinsic fitness. This behavior persists until values of $\mu$ beyond which selective adaptation is ineffective and the quasispecies collapse \cite{eigen1971selforganization},  analogous to the error catastrophe (cf. \emph{Supplementary Methods}). 

The presence of an immune response changes the ``effective" fitness landscape by favoring mutations that enable the virus to escape immune pressure despite lowering intrinsic fitness. This causes the dominant viral quasispecies to shift away from the WT strain, and the viral quasispecies sample different parts of sequence space in different hosts because of the great diversity of human genes associated with T cell immune responses (Fig.~\ref{fig2}(b) and Supplementary Fig. S9). Primary mutations that enable escape also influence the emergence of secondary mutations at sites which are not directly targeted by the immune pressure, but are coupled to the primary mutations to compensate the incurred fitness cost \cite{bonhoeffer2004evidence,dahirel2011coordinate,ferguson2012hivtrap}. At the population level, Fig.~\ref{fig2}(b) shows that due to immune pressure, the consensus sequences from different hosts explore diverse regions of the fitness landscape. Thus, the immune pressure imposed by different hosts acts like a ``higher temperature" that facilitates sampling of sequence space. 

\begin{figure*}[htp!]
\centering
\includegraphics[scale=0.63]{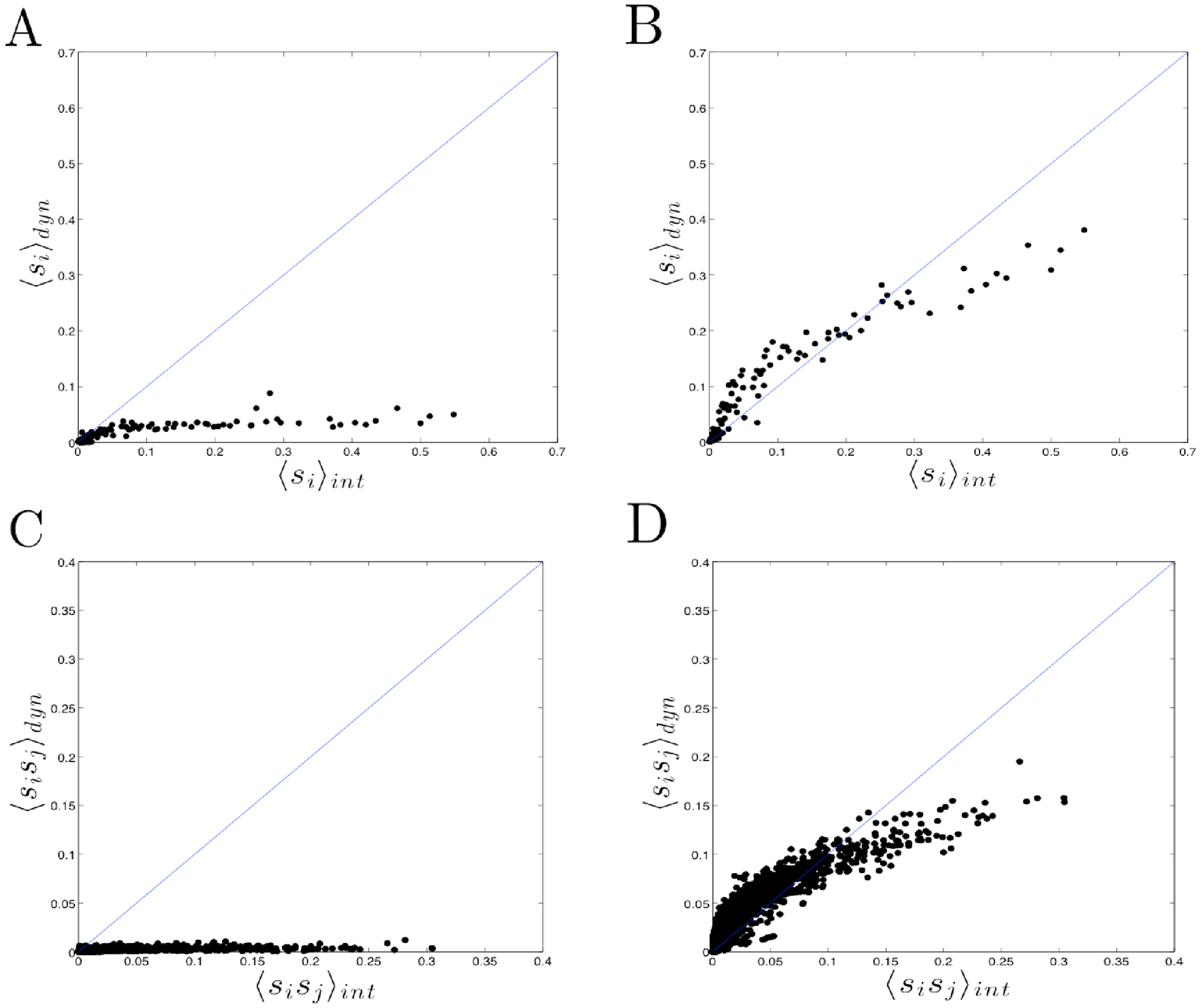}
\caption{\textbf{Comparison of one and two-body mutational probabilities at low and intermediate mutation rates {\color{red}}}\\
 (a), (c) $\mu$ = $5\times10^{-5}$/site/generation, $N_v$ = 15000. The one and two-body mutational probabilities computed from the population ensemble resulting from our simulations, $\langle s_i \rangle_{dyn}$ and $\langle s_i s_j \rangle_{dyn}$, are compared with their  counterparts, $\langle s_i \rangle_{int}$ and $\langle s_i s_j \rangle_{int}$, both computed from an equilibrium sampling of the intrinsic fitness Hamiltonian $H_{int}[\vec{s}\,]$, which agree with values computed from sequences used to infer the maximum entropy model. At small mutation rates, escape mutations are rarely sampled. The viral quasispecies within each host stays frozen near the ground state (cf. Fig. \ref{fig2}a) and mutations are not selected at the population level, resulting in $\langle s_i \rangle_{dyn}, \, \langle s_i s_j \rangle_{dyn} \approx 0$. (b), (d) For intermediate mutation rates $\mu \in (10^{-4}, 10^{-2})$ (here, $\mu = 5\times10^{-3}$/site/generation and $N_v$ = 15000), we find that immune selection leads to the accumulation of mutations across the viral proteome, and at the population level the one and two-body mutational probabilities $\langle s_i \rangle_{dyn}$ and $\langle s_i s_j \rangle_{dyn}$ correlate monotonically with their counterparts $\langle s_i \rangle_{int}$ and $\langle s_i s_j \rangle_{int}$. At higher mutation rates ($\mu$ $ > 10^{-2}$/site/generation), the quasispecies within most hosts become unable to survive selection as deleterious mutations are rapidly accumulated, analogous to the phenomenon of error catastrophe \cite{eigen1971selforganization}. }
\label{fig3}
\end{figure*}

The exploration of sequence space is sensitive to the value of the mutation rate $\mu$ (cf. Fig. \ref{fig3}). There is an intermediate range $\mu \in (10^{-4}, 10^{-2})$ where immune selection is stable and favors viral adaptation. This range is higher than the mutation rate of HIV ($\sim 10^{-4}$) when measured in units of per amino acid site per replication cycle.  But, this is reasonable as each generation in our simulations corresponds to a number of replication cycles for reasons described earlier, and we are simulating a single protein with only $\sim 100$ sites as compared to the whole HIV proteome ($\sim 3000$ sites).  

In this intermediate range of $\mu$, the marginal single, two and three site probabilities $\langle s_i \rangle_{dyn}$, $\langle s_i s_j \rangle_{dyn}$ and  $\langle s_i s_j s_k \rangle_{dyn}$ are monotonically correlated  with their intrinsic fitness counterparts $\langle s_i \rangle_{int}$, $\langle s_i s_j \rangle_{int}$ and $\langle s_i s_j s_k \rangle_{int}$ (Fig.~\ref{fig3}(b), (d) and Supplementary Fig. S10). Taken together, our results suggest that immune pressure plays a necessary role in facilitating exploration of sequence space so that the viral quasispecies sample the fitness landscape, and furthermore, the correlation structure of the prevailing consensus strains in the population ensemble is monotonically related to the correlations that characterize mutant strains selected according to intrinsic replicative fitness. 

\section{Variational theory}

Our ultimate interest, however, is not just in correlation structure, but in characterizing the relationship between the intrinsic fitness landscape of the virus, and the prevalence landscape inferred from patient-derived sequences. Toward this goal we exploit a mapping by Leuth\"{a}usser~\cite{leuthausser1987eigen} to describe non-equilibrium quasispecies evolution according to Eigen's equation~\cite{eigen1971selforganization}:
Each evolutionary path in sequence space is denoted by 
$\Sigma = \{ \vec{s^0},\,\vec{s^1},\,\vec{s^{2}},\,\ldots \vec{s^n } \}$, 
where $\vec{s^{\alpha}}$ denotes a strain in generation $\alpha$, and can be regarded as a configuration of an inhomogenous Ising model. Different generations $\vec{s^{\alpha}}$ realized in a particular evolutionary path correspond to sequentially arranged rows of this Ising system. 
The probability of a particular evolutionary path, $\Sigma$, is 
$p(\Sigma) \propto e^{-\mathcal{H}(\Sigma)}$ with the Hamiltonian 
\begin{equation}
\mathcal{H}(\Sigma) = -J  \sum_{\alpha} \left(\vec{1} - 2\vec{s^{\alpha}}  \right). \left( \vec{1} - 2\vec{s^{\alpha+1}} \right) + \sum_{\alpha} H^{\alpha} [\vec{s^{\alpha}}] \,.
\label{leuthausser_H}
\end{equation}
The first term in equation (\ref{leuthausser_H}) describes a coupling between the same site in successive   generations, with $J = \frac{1}{2}\log \left(\frac{1-\mu}{\mu} \right)$ (since $\mu$ is small, $J$ is positive, preferring sites in successive generations to be the same \cite{leuthausser1987eigen}). This describes the phylogenetic relationship between strains in a population. The second term in $\mathcal{H}(\Sigma)$ reflects the effective fitness of a particular strain, which we decompose as 
\begin{eqnarray}
H^{\alpha} [ \vec{s^{\alpha}}]  &=& H_{int}[ \vec{s^{\alpha}}] + I^{\alpha}[ \vec{s^{\alpha}}] \nonumber \\
	& =&  \displaystyle \sum_{i < j=1}^N J_{ij} s_i^{\alpha} s_j^{\alpha} + \displaystyle \sum_{i=1}^N h_i s_i^{\alpha} - \displaystyle \sum_i b_i^{\alpha} s_i^{\alpha}
\label{leuthausser_H2}
\end{eqnarray}
$H_{int}$ (parameterized by $\{ J_{ij},\, h_i\}$) is $\alpha$ independent, while the immune pressure in generation $\alpha$ is described by the fields $\{b_i^{\alpha}\}$. 

The non-equilibrium dynamics captured by equations (\ref{leuthausser_H}, \ref{leuthausser_H2}), while not identical to our simulations, contain the important elements of phylogeny and immune pressure in different hosts. Instead of numerically sampling configurations with normalized probability $p(\Sigma) =e^{-\mathcal{H}(\Sigma)}/Z$ (which we expect will lead to results consistent with our simulations), we developed analytical approximations. The prevalence landscape in equation (\ref{maxent}), inferred from sequences extracted from patients at a multitude of different times, makes no reference to phylogeny or immune pressure. Thus, we asked how well a ``phylogeny and immune pressure independent"  probability of the form $p_T(\Sigma)=(\prod_\alpha e^{-H_T^{\alpha}})/Z_T$ can approximate $p(\Sigma)$.

In particular, we chose a trial Hamiltonian of the same form as the inferred Hamiltonian in equation (\ref{maxent}), $ H_T^{\alpha} (\{ K_{ij}^\alpha, a_i^\alpha \}, \vec{s^{\alpha}}) = \sum_{i < j = 1}^N K_{ij}^\alpha s_i^{\alpha} s_j^{\alpha} +  \sum_i a_i^\alpha s_i^{\alpha}$, and approximate $\mathcal{H}(\Sigma)$ in equation (\ref{leuthausser_H}) as  $\mathcal{H}_T = \sum_{\alpha} H_T^{\alpha} (\{ K_{ij}^\alpha, a_i^\alpha \}, \vec{s^{\alpha}})$. We then variationally estimate the $\alpha$-dependent parameters $\{K_{ij}^{\alpha}, a_i^{\alpha}\}$ which best approximate equation (\ref{leuthausser_H}).
These parameters can be estimated through  the Gibbs-Feynman-Bogoliobov variational bound~\cite{feynman1965quantum}  

\begin{equation}
\ln Z \geq \ln Z_T - \langle \mathcal{H} - \mathcal{H}_T \rangle_T.
\label{gibbs}
\end{equation}

Extremizing the bound through variations of the parameters $\{K^\alpha _{ij}\}$ and $\{a^\alpha _i\}$ leads to the self-consistent mean-field relations (cf. Appendix \ref{appD}),
\begin{eqnarray}
 a^{\alpha}_i &=& h_i  +4J\left(1-\langle s^{\alpha+1}_i \rangle_T-\langle s^{\alpha-1}_i \rangle_T\right) -{b^\alpha_i} \label{eq_var1}\\
 K^{\alpha}_{ij} &=&   J_{ij} \; \label{eq_var2}
 \end{eqnarray}
 Note that the effect of phylogeny and immune pressure appears only through the onsite fields.

Within the variational approximation, the probability of a strain $\vec{s}$ at a particular generation $\alpha$ is determined by $H_T^{\alpha} (\vec{s})$ and is independent of other generations. In contrast, the prevalence of $\vec{s}$ as encoded by the Hamiltonian in equation (\ref{maxent}) is inferred from data taken from different hosts in whom the virus has been sampled at different times. This is mimicked by averaging  $H_T^{\alpha} (\vec{s})$ over a large number of generations, leading to
\begin{equation}
 H_T[\vec{s}\,] = \displaystyle \sum_{i<j = 1}^N J_{ij} s_i s_j + \displaystyle \sum_{i=1}^N 
\left[h_i  +4J\left(1-2\langle s_i \rangle_T\right) - \bar{b_i}\right] s_i\;
 \label{averagedMF}
\end{equation}
where $\bar{b_i} = \frac{1}{n} \sum_{\alpha} b_i^{\alpha}$ is the average immune pressure,
and $\langle s_i \rangle_T$, the average single-site mutational probability, at site $i$.

Comparing the parameters in $H_T[\vec{s}\,]$ to that in $H_{int}[\vec{s}\,]$, we see that while the coupling constant between sites remains unmodified, the effective field at a particular site is changed from $h_i$ by two competing contributions. The first term is self-consistently related to $\langle s_i \rangle_T$, the average single-site mutation probability at site $i$ under Hamiltonian $H_T[\vec{s}\,]$. For $\langle s_i \rangle_T < 1/2$, which is true for $J>0$, the first term represents an increase in the field at site $i$ that disfavors mutation. This is because replicative fidelity disfavors sampling of sequence space since one mutant strain must be the progeny of another one.

In the absence of immune pressure, this ``phylogenetic coupling" term favors freezing into the ground state ($\vec{s}=\vec{0}$) accounting for the localization of quasispecies in the vicinity of the intrinsically fittest sequence. The immune response counters this effect, bringing the effective field $a_i$ closer to the intrinsic field $h_i$, and drives the statistics of mutations closer to those governed by intrinsic fitness.  

Armed with this variational approximation, we can now ask if $H_T[\vec{s}\,]$ preserves the fitness ranks of different viral strains as encoded in $H_{int}[\vec{s}\,]$. The values of $\{\bar{b_i}\}$ are easily obtained from the simulations; and $\langle s_i \rangle_T$ are approximated by $\langle s_i \rangle_{dyn}$ from our simulations (calculating $\langle s_i \rangle_T$ self-consistently is not practical given the complexity of $H_T$). This assumes that the variational Hamiltonian $\mathcal{H}_T$ is a reasonable approximation of true quasispecies evolution according to Eigen's equation, and that the latter (where the number of strains is unbounded) is a reasonable facsimile  of our simulations (where the quasispecies population is bounded). The second claim is supported by Dixit et al. \cite{dixit2012finite} who formally prove for a class of finite population evolution models (similar to the one we have considered in this work) that as the population size increases, the stationary distribution of genotypes converges to the distribution predicted by Eigen's quasispecies model \cite{dixit2012note}.  

\begin{figure}[htp!]
\centering
\includegraphics[scale=0.5]{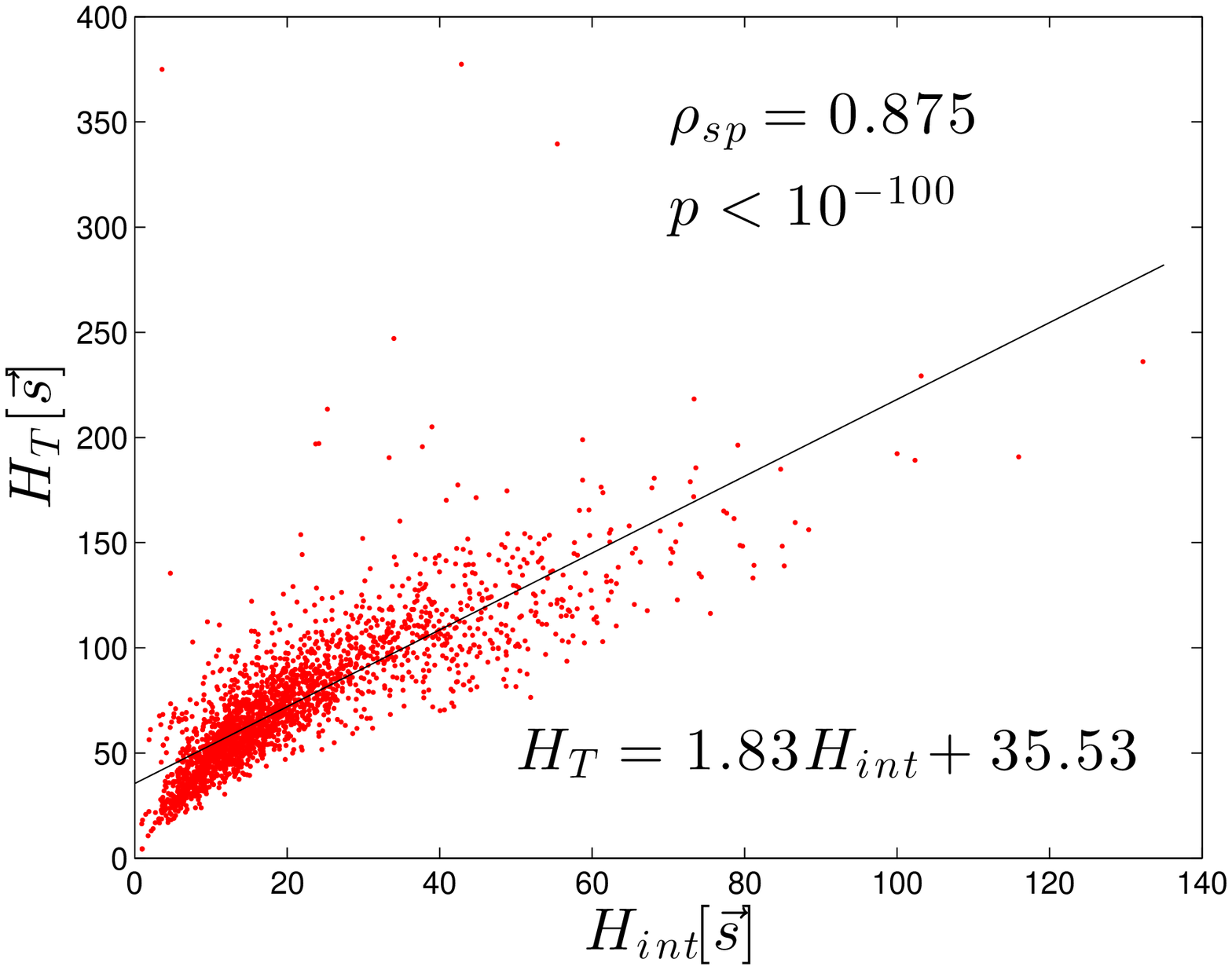}
\caption{\textbf{Variational analysis using the Feynman bound predicts that the prevalence landscape is correlated with the fitness landscape}\\
A numerical comparison between the variational estimate of the prevalence Hamiltonian $H_T[\vec{s}\,]$ (equation (\ref{averagedMF})) to the intrinsic Hamiltonian $H_{int}[\vec{s}\,]$ for the case for 2500 subtype B p17 sequences. The parameters $\mu = 5\times10^{-3}$/site/generation, $n_{max} = 6$, $N_v=15000$ were used in the simulations. Each point on the plot corresponds to one p17 sequence. The sequences were downloaded from the Los Alamos sequence database \cite{hivdb} and converted to the binary code. $\rho_{sp}$ is the standard rank correlation computed from Spearman's test \cite{corder2009nonparametric} and $p$ is the corresponding significance value. The line is computed from fitting an ordinary least squares regression model.}
\label{fig4}
\end{figure}

In Fig.~\ref{fig4}, we plot $H_T[\vec{s}\,]$ versus $H_{int}[\vec{s}\,]$ for 2474 subtype B HIV-1 strains extracted from a public database \cite{hivdb}, after converting them to the binary code. A Spearman rank test \cite{corder2009nonparametric} shows that the order of ranking is preserved with very high statistical accuracy for most strains ($\rho = 0.875$, $p < 10^{-100}$). Thus, at least within a mean-field approximation, prevalence landscapes inferred from patient-derived virus protein sequences preserve the rank-order of intrinsic replicative fitnesses of mutant virus strains. 

\section{Discussion}

The underlying reason for this result may be simple.  Because of the great diversity of genes that determine the immune response, individual sites in viral proteins are targeted by a small fraction of infected patients (see \emph{Supplementary Methods}).  Furthermore, clinical data show that when the immune pressure in a particular host results in escape mutations, and the mutated virus is transmitted to another host who does not target the mutated sites, the virus rapidly reverts to WT in these regions~\cite{henn2012whole}. Within a given host, the magnitude of immune pressure at particular sites ($b_i^{\alpha}$) is large enough to drive exploration of sequence space. But this effect is present at any site only in a small fraction of hosts, and acts as a perturbation ($\bar{b_i}$) when averaged over many consensus sequences.  Therefore, although the immune pressure imposed by genetically diverse patients enables exploration of sequence space by modifying the fitness landscape (Fig.~\ref{fig2}(b)), using a sufficient number of sequences ensures that the inferred prevalence model preserves the rank order of the intrinsic fitnesses of mutant viral strains.   

This should, however, only be true if we compare sequences that are not phylogenetically distant. The effects of phylogeny, immune pressure, and intrinsic fitness are concatenated in the parameters that define our inferred fitness landscape (equation (\ref{averagedMF})). For reasons discussed above, the immune pressure is a critical, but perturbative field. The effects of phylogeny can be strong, but are similar for phylogenetically related strains. This effect can, however, be quite different for phylogenetically distal strains. In other words, because of replicative fidelity, a phylogenetically distant strain is less likely to be prevalent than one that is phylogenetically closer, even if they are of comparable fitness.  The excellent agreement between experimental measurements of replicative fitness and the inferred prevalence landscape described by Ferguson et al. \cite{ferguson2012hivtrap} may reflect the fact that different strains were phylogenetically proximal.  This is also likely the case for strains used to construct Fig. \ref{fig4}.  From a practical standpoint of using the fitness landscape for immunogen design, this issue presents little difficulty as a vaccine-induced immune response is unlikely to generate mutants that are phylogenetically distal.  But, care needs to be taken in using the inferred prevalence landscape when comparing in vitro fitness measurements of phylogenetically distant strains. This is because our estimate of the correction due to replicative fidelity/phylogeny in equation (\ref{averagedMF}) is not expected to be quantitatively correct. 

The availability of fitness landscapes of viruses can accelerate the rational design of immunogens that may be able to induce potent and effective immune responses that protect humans from infectious diseases.  Taken together, our results, and those in Ferguson et al. \cite{ferguson2012hivtrap} show that maximum entropy models inferred from viral protein sequences sampled from patients can faithfully represent the intrinsic fitness landscape for phylogenetically related strains. Further work needs to be done to develop general methods for deconvoluting the effects of phylogeny and intrinsic fitness in inferred landscapes in order to reliably predict the fitness of strains regardless of phylogenetic distance. With the rapid expansion of available genomic data a promising and efficient route to rational immunogen design is thus suggested.

\section*{Acknowledgments}
This research was supported by the Ragon Institute of MGH, MIT, and Harvard, and a NIH DirectorÕs Pioneer award (AKC). Fruitful discussions with Dr. T. Butler are gratefully acknowledged.

\renewcommand{\theequation}{A\@arabic\c@equation} 
\section*{Appendix}  
\appendix

\section{Modeling immune pressure}
%\hypertarget{appA}{}
\label{appA}
The immune response in a particular host is chosen to randomly target $k$ sites in the protein where $k$ is a random integer between 0 and $n_{max} \ll N$, based on clinical evidence within Caucasian Americans that the p17 protein is targeted by T cells, and that a given protein site is expected to be targeted by a very small fraction of individuals in a population (\emph{Supplementary Methods} and Supplementary Fig. S5). For the simulations reported in the main text, we employ $n_{max}=6$. The targeted sites $\alpha_1,\,\alpha_2,\,\ldots,\,\alpha_k$ are chosen randomly from protein sites $1,\,2,\,\ldots,\,N$ in each host without bias, thereby mimicking the highly polymorphic nature of genes that encode the cell machinery that presents viral protein fragments that are recognized by T cells. Thus, each individual is likely to target different sites compared to other persons. We assume that the parameters $\{b_{\alpha_i}\}_{i=1}^k$ are independent random variables,
drawn from the same Gaussian distribution of mean $\bar{h}$ and variance $\sigma_h^2$ as determined by the intrinsic fitness parameters $\{h_i\}$. Changing the parameters of this distribution to increase the magnitude of the immune fields does not change the main conclusions of our work (see \emph{Supplementary Methods}) as long as the immune pressure in any given host targets only a small fraction of sites ($<$10\%) and that the typical magnitude of $\{b_i\}$ do not greatly exceed $\bar{h}$. 

\section{Quasispecies simulations}
\label{appB}
Within each host (parameterized by immune pressure $\{b_i\}$) the viral quasispecies evolve for $\tau_S$ generations following infection with $N_v$  copies of a ``founder" strain. Each generation is comprised of the following steps,
\begin{enumerate}
\item{\textit{Mutation: } For each viral strain $\vec{s}$ within the quasispecies, every site $i$ is mutated with probability $\mu$. In the binary representation, this amounts to the operation $s_i \rightarrow 1-s_i$}. 
\item{\textit{Pre-screening:} Eliminate sequences that escape the region in the reduced free energy landscape defined by subtype B sequences (see \emph{Supplementary Methods} and Supplementary Fig. S2)}. (This step is rarely necessary unless the mutation rate $\mu$ exceeds $10^{-2}$).
\item{\textit{Selection:} Strain $\vec{s}$ is selected to survive with probability $p_s(\vec{s}\,) = \frac{e^{-H[\vec{s}\,]}}{1+e^{-H[\vec{s}\,]}}$. Here, $H[\vec{s}\,] = H_{int}[\vec{s}\,]- \sum_{i=1}^N b_i s_i$ and parameters $\{b_i\}$ are host-specific.}
\item{\textit{Replenishment:} The numbers of surviving strains is randomly resampled with replacement to replenish the viral population to size $N_v$.}
\end{enumerate}

The survival probability in step 3 above has a functional form consistent to the ``death probability" employed in Amitrano et al. \cite{amitrano1989evol}. Assuming $f_{\vec{s}} \sim e^{-H[\vec{s}\,]}$ is the fitness of strain $\vec{s}$, the survival probability has a simple interpretation. In each generation a strain $\vec{s}$ is compared with a WT strain and is elected to survive with probability $\frac{f_{\vec{s}}}{f_{WT} + f_{\vec{s}}}$, where $f_{WT} \sim e^{-H_{WT}} = 1$.  Alternatively, one can envision a selection rule where a strain $\vec{s}$ is compared with the average strain in the current quasispecies and is elected to survive with probability $\frac{f_{\vec{s}}}{\bar{f_{\vec{s}}} + f_{\vec{s}}}$, where $\bar{f_{\vec{s}}} \sim \langle e^{-H[\vec{s}\,]}\rangle$ is the average fitness of the quasispecies. The latter selection rule produces results that are in qualitative agreement with the former selection rule within the range of parameters we have explored (data not shown).

\section{Landscape visualization}
\label{appC}
Mutational states of the protein were sampled according to $H_{int}[\vec{s}\,]$ by the Metropolis Monte-Carlo (MC) algorithm~\cite{metropolis1953equation} to generate an ``equilibrium" ensemble of $\approx 10^6$ sequences. We applied principal component analysis (PCA)~\cite{friedman2001elements} to the covariance matrix corresponding to double mutations in this ensemble of sequences (see \emph{Supplementary Methods}). We computed the projection of each sequence in the equilibrium ensemble onto this space along the top two principal components (PCs). Using an appropriately sized square mesh, the density of sequences at different locations in this 2-D embedding was converted to the analog of free energy contours in statistical mechanics using the relation,
\begin{equation}
A(x,y) = - \log P(x,y)  \nonumber
\end{equation}
where $(x, y)$ is the center of a cell in the $[PC_1, \,PC_2]$ plane and $P(x,\,y)$ is the sample probability of a sequence in the equilibrium ensemble occupying this point. 

In the projection along $PC_1-PC_2$, the landscape exhibits three high fitness (or low ``free energy'') peaks (see \emph{Supplementary Methods} and Supplementary Fig. S2). But two of these peaks are unexplored by  the subtype B sequences in the MSA that were used to parametrize $H_{int}[\vec{s}\,]$ (Supplementary Fig. S3) and represent extrapolations of the fitted non-linear model. To focus on the question of how the inferred prevalence of subtype B HIV strains in the population relates to the intrinsic fitness landscape, we restricted quasispecies sequences in our computer simulations to lie in the region corresponding to observed sequences by placing reflecting boundaries on the $PC_1-PC_2$ space as described in Supplementary Fig. S2. 

\section{Variational calculations}
\label{appD}
In variational mean-field theory, the parameters of the approximate Hamiltonian are obtained by maximizing the RHS of the Gibbs-Feynman-Bogoliobov bound \cite{feynman1965quantum} (equation (4)) with respect to the variational parameters $\{K_{ij}^{\alpha}, \, a_i^{\alpha}\}$ . The stationarity conditions are,

\begin{align}
  \label{stationarity}
  \begin{split}
 \frac{\partial \ln Z_T}{\partial K_{ij}^{\alpha}} - \frac{\partial}{\partial K_{ij}^{\alpha}} \langle \mathcal{H} - \mathcal{H}_T \rangle_T \; =& \; 0\\
\frac{\partial \ln Z_T}{\partial a_{i}^{\alpha}} - \frac{\partial}{\partial a_{i}^{\alpha}} \langle \mathcal{H} - \mathcal{H}_T \rangle_T \; =& \; 0 
  \end{split}
\end{align}

Here, $\mathcal{H}$ is the original Hamiltonian while $\mathcal{H}_T = \sum_{\alpha} H_T^{\alpha} (\{ K_{ij}^\alpha, a_i^\alpha \}, \vec{s^{\alpha}})$ is the trial Hamiltonian with the form, $ H_T^{\alpha} (\{ K_{ij}^\alpha, a_i^\alpha \}, \vec{s^{\alpha}}) =  \sum_{i < j = 1}^N K_{ij}^\alpha s_i^{\alpha} s_j^{\alpha} +  \sum_i a_i^\alpha s_i^{\alpha}$. The trial partition function is defined as $Z_T  = \sum_{\{\vec{s^{\alpha}} \} }  \prod_{\alpha} \exp \left( - H_T^{\alpha} \left[ \{ K_{ij}^{\alpha}, a_i^{\alpha} \}, \vec{s^{\alpha}} \right] \right)$. As different generations are uncoupled from each other the sum and the product can be interchanged. Taking its logarithm,

\begin{eqnarray}
\ln Z_T &=& \displaystyle \sum_{\alpha} \ln \left( \displaystyle \sum_{\vec{s^{\alpha}}} \exp \left( - H_T \left[ \{ \Kij, \ai \}, \vec{s^{\alpha}} \right]\right) \right) \nonumber \\
             &=& \displaystyle \sum_{\alpha} \ln Z_T^{\alpha}
\end{eqnarray}

where $Z_T^{\alpha}$ is the partition function for a single generation in the evolutionary trajectory. 

Thus, Eq. \ref{stationarity} can be further simplified as,

\begin{align}
  \label{stationarity1}
  \begin{split}
 \frac{\partial \ln Z_T^{\alpha}}{\partial K_{ij}^{\alpha}} - \frac{\partial}{\partial K_{ij}^{\alpha}} \langle \mathcal{H} - \mathcal{H}_T \rangle_T \; =& \; 0\\
\frac{\partial \ln Z_T^{\alpha}}{\partial a_{i}^{\alpha}} - \frac{\partial}{\partial a_{i}^{\alpha}} \langle \mathcal{H} - \mathcal{H}_T \rangle_T \; =& \; 0 
  \end{split}
\end{align}

As is well-known in equlibrium statistical mechanics, $\ln Z_T^{\alpha}$ is the equivalent of a scaled free energy and its derivatives with regards to the coupling constants and fields yield thermal averages of different quantities of interest. Thus, it can be shown that $\frac{\pr \ln Z_T^{\alpha}}{\pr K_{ij}^{\alpha}} = - \langle s_i^{\alpha} s_j^{\alpha} \rangle_T$ and  $\frac{\pr \ln Z_T^{\alpha}}{\pr a_{i}^{\alpha}} = - \langle s_i^{\alpha}\rangle_T$. Substituting in equation (\ref{stationarity1}),

\begin{align}
\label{var1}
\begin{split}
- \langle s_i^{\alpha} s_j^{\alpha} \rangle_T - \frac{\pr \langle \mathcal{H} \rangle_T}{\pr \Kij} +  \frac{\pr \langle \mathcal{H}_T \rangle_T}{\pr \Kij} \; =& \; 0 \\
- \langle s_i^{\alpha} \rangle_T - \frac{\pr \langle \mathcal{H} \rangle_T}{\pr \ai} +  \frac{\pr \langle \mathcal{H}_T \rangle_T}{\pr \ai}  \; =& \; 0
\end{split}
\end{align}

Substituting expressions for $\langle \mathcal{H} \rangle_T$ (cf. main text) and $\langle \mathcal{H}_T \rangle_T$,

\begin{widetext}
\begin{align}
\label{var1}
\begin{split}
- \cancelto{}{\langle s_i^{\alpha} s_j^{\alpha} \rangle_T} - \displaystyle \sum_{k < m, \beta} J_{km} \frac{\pr  \langle s_k^{\beta} s_m^{\beta} \rangle_T  }{\pr \Kij} -  \s_{k, \beta} (h_k - b_k^{\beta}) \frac{\pr \langle s_k^{\beta} \rangle_T}{\Kij} - 2J \s_{k, \beta} \frac{\pr }{ \pr \Kij} \left( \langle s_k^{\beta} \rangle_T + \langle s_k^{\beta + 1} \rangle_T \right)  & \\
+\: \: 4J \frac{\pr \langle s_k^{\beta} s_k^{\beta+1} \rangle_T}{\Kij} + \cancelto{}{\langle s_i^{\alpha} s_j^{\alpha} \rangle_T} + \s_{k < m, \beta} K_{km}^{\beta} \frac{\pr  \langle s_k^{\beta} s_m^{\beta} \rangle_T  }{\pr \Kij} + \s_{k, \beta} a_k^{\beta} \frac{\pr \langle s_k^{\beta} \rangle_T}{\Kij} \; =& \; 0 \\
- \cancelto{}{\langle s_i^{\alpha} \rangle_T} - \displaystyle \sum_{k < m, \beta} J_{km} \frac{\pr  \langle s_k^{\beta} s_m^{\beta} \rangle_T  }{\pr \ai} -  \s_{k, \beta} (h_k - b_k^{\beta}) \frac{\pr \langle s_k^{\beta} \rangle_T}{\ai} - 2J \s_{k, \beta} \frac{\pr }{ \pr \ai} \left( \langle s_k^{\beta} \rangle_T + \langle s_k^{\beta + 1} \rangle_T \right)  \\
+\: \: 4J \frac{\pr \langle s_k^{\beta} s_k^{\beta+1} \rangle_T}{\ai} + \cancelto{}{\langle s_i^{\alpha} \rangle_T} + \s_{k < m, \beta} K_{km}^{\beta} \frac{\pr  \langle s_k^{\beta} s_m^{\beta} \rangle_T  }{\pr \ai} + \s_{k, \beta} a_k^{\beta} \frac{\pr \langle s_k^{\beta} \rangle_T}{\ai} \; =& \; 0
\end{split}
\end{align}
\end{widetext}

The derivative encoding the phylogenetic (inter-generational) coupling can be simplified as follows,

\begin{align}
\begin{split}
\s_{k, \beta}  \frac{\pr \langle s_k^{\beta} s_k^{\beta+1} \rangle_T}{\Kij} =& \s_{k, \beta} \left( \langle s_k^{\beta} \rangle_T \frac{\pr \langle s_k^{\beta + 1} \rangle_T}{\pr \Kij} +  \langle s_k^{\beta+1} \rangle_T \frac{\pr \langle s_k^{\beta} \rangle_T}{\pr \Kij} \right) \\
=& \s_{k, \beta} \left( \langle s_k^{\beta-1} \rangle_T + \langle s_k^{\beta+1} \rangle_T \right) \frac{\pr \langle s_k^{\beta} \rangle_T}{\pr \Kij}
\end{split}
\end{align}

and similarly for the derivative with respect to $\ai$. This simplifies the variational equations to,

\begin{widetext}
\begin{align}
\label{var_final}
\begin{split}
\s_{k < m, \beta}\left\{\Kkm - J_{km} \right\} \frac{\pr  \langle s_k^{\beta} s_m^{\beta} \rangle_T  }{\pr \Kij} + \s_{k, \beta} \left\{ \ak -h_k  + b_k^{\beta} - 4J\left( 1 - \langle s_k^{\beta-1} \rangle_T - \langle s_k^{\beta+1} \rangle_T \right)  \right\}  \frac{\pr \langle s_k^{\beta} \rangle_T}{\Kij} =& \; 0 \\
\s_{k < m, \beta}\left\{\Kkm - J_{km} \right\} \frac{\pr  \langle s_k^{\beta} s_m^{\beta} \rangle_T  }{\pr \ai} + \s_{k, \beta} \left\{ \ak -h_k  + b_k^{\beta} - 4J\left( 1 - \langle s_k^{\beta-1} \rangle_T - \langle s_k^{\beta+1} \rangle_T \right)  \right\} \frac{\pr \langle s_k^{\beta} \rangle_T}{\ai} \; =& \; 0
\end{split}
\end{align}
\end{widetext}

Eqs. \ref{var_final} are satisfied if the coefficients of the derivatives are set identically to zero and we obtain equations (\ref{eq_var1}) and (\ref{eq_var2}).

\bibliography{biblio2.bib}

\begin{thebibliography}{29}%
\makeatletter
\providecommand \@ifxundefined [1]{%
 \@ifx{#1\undefined}
}%
\providecommand \@ifnum [1]{%
 \ifnum #1\expandafter \@firstoftwo
 \else \expandafter \@secondoftwo
 \fi
}%
\providecommand \@ifx [1]{%
 \ifx #1\expandafter \@firstoftwo
 \else \expandafter \@secondoftwo
 \fi
}%
\providecommand \natexlab [1]{#1}%
\providecommand \enquote  [1]{``#1''}%
\providecommand \bibnamefont  [1]{#1}%
\providecommand \bibfnamefont [1]{#1}%
\providecommand \citenamefont [1]{#1}%
\providecommand \href@noop [0]{\@secondoftwo}%
\providecommand \href [0]{\begingroup \@sanitize@url \@href}%
\providecommand \@href[1]{\@@startlink{#1}\@@href}%
\providecommand \@@href[1]{\endgroup#1\@@endlink}%
\providecommand \@sanitize@url [0]{\catcode `\\12\catcode `\$12\catcode
  `\&12\catcode `\#12\catcode `\^12\catcode `\_12\catcode `\%12\relax}%
\providecommand \@@startlink[1]{}%
\providecommand \@@endlink[0]{}%
\providecommand \url  [0]{\begingroup\@sanitize@url \@url }%
\providecommand \@url [1]{\endgroup\@href {#1}{\urlprefix }}%
\providecommand \urlprefix  [0]{URL }%
\providecommand \Eprint [0]{\href }%
\providecommand \doibase [0]{http://dx.doi.org/}%
\providecommand \selectlanguage [0]{\@gobble}%
\providecommand \bibinfo  [0]{\@secondoftwo}%
\providecommand \bibfield  [0]{\@secondoftwo}%
\providecommand \translation [1]{[#1]}%
\providecommand \BibitemOpen [0]{}%
\providecommand \bibitemStop [0]{}%
\providecommand \bibitemNoStop [0]{.\EOS\space}%
\providecommand \EOS [0]{\spacefactor3000\relax}%
\providecommand \BibitemShut  [1]{\csname bibitem#1\endcsname}%
\let\auto@bib@innerbib\@empty
%</preamble>
\bibitem [{\citenamefont {Gaschen}\ \emph {et~al.}(2002)\citenamefont {Gaschen}
  \emph {et~al.}}]{gaschen2002diversity}%
  \BibitemOpen
  \bibfield  {author} {\bibinfo {author} {\bibfnamefont {B.}~\bibnamefont
  {Gaschen}} \emph {et~al.},\ }\bibfield  {title} {\enquote {\bibinfo {title}
  {Diversity considerations in {HIV}-1 vaccine selection},}\ }\href@noop {}
  {\bibfield  {journal} {\bibinfo  {journal} {\emph{Science}}\ }\textbf
  {\bibinfo {volume} {296}},\ \bibinfo {pages} {2354--2360} (\bibinfo {year}
  {2002})}\BibitemShut {NoStop}%
\bibitem [{\citenamefont {Goulder}\ and\ \citenamefont
  {Watkins}(2004)}]{goulder2004escape}%
  \BibitemOpen
  \bibfield  {author} {\bibinfo {author} {\bibfnamefont {P.~J.~R.}\
  \bibnamefont {Goulder}}\ and\ \bibinfo {author} {\bibfnamefont {D.~I.}\
  \bibnamefont {Watkins}},\ }\bibfield  {title} {\enquote {\bibinfo {title}
  {{HIV} and {SIV} {CTL} escape: {I}mplications for vaccine design},}\
  }\href@noop {} {\bibfield  {journal} {\bibinfo  {journal} {\emph{Nature
  Reviews Immunology}}\ }\textbf {\bibinfo {volume} {4}},\ \bibinfo {pages}
  {630--640} (\bibinfo {year} {2004})}\BibitemShut {NoStop}%
\bibitem [{\citenamefont {Walker}\ and\ \citenamefont
  {Burton}(2008)}]{walker2008vaccine}%
  \BibitemOpen
  \bibfield  {author} {\bibinfo {author} {\bibfnamefont {B.~D.}\ \bibnamefont
  {Walker}}\ and\ \bibinfo {author} {\bibfnamefont {D.~R.}\ \bibnamefont
  {Burton}},\ }\bibfield  {title} {\enquote {\bibinfo {title} {Toward an {AIDS}
  vaccine},}\ }\href@noop {} {\bibfield  {journal} {\bibinfo  {journal}
  {\emph{Science}}\ }\textbf {\bibinfo {volume} {320}},\ \bibinfo {pages}
  {760--764} (\bibinfo {year} {2008})}\BibitemShut {NoStop}%
\bibitem [{\citenamefont {L{\'e}tourneau}\ \emph {et~al.}(2007)\citenamefont
  {L{\'e}tourneau} \emph {et~al.}}]{letourneau2007vaccine}%
  \BibitemOpen
  \bibfield  {author} {\bibinfo {author} {\bibfnamefont {S.}~\bibnamefont
  {L{\'e}tourneau}} \emph {et~al.},\ }\bibfield  {title} {\enquote {\bibinfo
  {title} {Design and pre-clinical evaluation of a universal {HIV-1}
  vaccine},}\ }\href@noop {} {\bibfield  {journal} {\bibinfo  {journal}
  {\emph{PLoS one}}\ }\textbf {\bibinfo {volume} {2}},\ \bibinfo {pages} {e984}
  (\bibinfo {year} {2007})}\BibitemShut {NoStop}%
\bibitem [{\citenamefont {Schneidewind}\ \emph {et~al.}(2007)\citenamefont
  {Schneidewind} \emph {et~al.}}]{schneidewind2007escape}%
  \BibitemOpen
  \bibfield  {author} {\bibinfo {author} {\bibfnamefont {A.}~\bibnamefont
  {Schneidewind}} \emph {et~al.},\ }\bibfield  {title} {\enquote {\bibinfo
  {title} {Escape from the dominant {HLA-B27}-restricted cytotoxic
  {T}-lymphocyte response in {Gag} is associated with a dramatic reduction in
  human immunodeficiency virus type 1 replication},}\ }\href@noop {} {\bibfield
   {journal} {\bibinfo  {journal} {\emph{Journal of Virology}}\ }\textbf
  {\bibinfo {volume} {81}},\ \bibinfo {pages} {12382--12393} (\bibinfo {year}
  {2007})}\BibitemShut {NoStop}%
\bibitem [{\citenamefont {Troyer}\ \emph {et~al.}(2009)\citenamefont {Troyer}
  \emph {et~al.}}]{troyer2009variable}%
  \BibitemOpen
  \bibfield  {author} {\bibinfo {author} {\bibfnamefont {R.~M.}\ \bibnamefont
  {Troyer}} \emph {et~al.},\ }\bibfield  {title} {\enquote {\bibinfo {title}
  {Variable fitness impact of {HIV-1} escape mutations to cytotoxic {T}
  lymphocyte ({CTL}) response},}\ }\href@noop {} {\bibfield  {journal}
  {\bibinfo  {journal} {\emph{PLoS pathogens}}\ }\textbf {\bibinfo {volume}
  {5}},\ \bibinfo {pages} {e1000365} (\bibinfo {year} {2009})}\BibitemShut
  {NoStop}%
\bibitem [{\citenamefont {Dahirel}\ \emph {et~al.}(2011)\citenamefont {Dahirel}
  \emph {et~al.}}]{dahirel2011coordinate}%
  \BibitemOpen
  \bibfield  {author} {\bibinfo {author} {\bibfnamefont {V.}~\bibnamefont
  {Dahirel}} \emph {et~al.},\ }\bibfield  {title} {\enquote {\bibinfo {title}
  {Coordinate linkage of {HIV} evolution reveals regions of immunological
  vulnerability},}\ }\href@noop {} {\bibfield  {journal} {\bibinfo  {journal}
  {\emph{Proc. Natl. Acad. Sci.}}\ }\textbf {\bibinfo {volume} {108}},\
  \bibinfo {pages} {11530--11535} (\bibinfo {year} {2011})}\BibitemShut
  {NoStop}%
\bibitem [{\citenamefont {Kouyos}\ \emph {et~al.}(2012)\citenamefont {Kouyos}
  \emph {et~al.}}]{kouyos2012landscape}%
  \BibitemOpen
  \bibfield  {author} {\bibinfo {author} {\bibfnamefont {R.~D.}\ \bibnamefont
  {Kouyos}} \emph {et~al.},\ }\bibfield  {title} {\enquote {\bibinfo {title}
  {Exploring the complexity of the {HIV-1} fitness landscape},}\ }\href@noop {}
  {\bibfield  {journal} {\bibinfo  {journal} {\emph{PLoS Genetics}}\ }\textbf
  {\bibinfo {volume} {8}},\ \bibinfo {pages} {e1002551} (\bibinfo {year}
  {2012})}\BibitemShut {NoStop}%
\bibitem [{\citenamefont {Ferguson}\ \emph {et~al.}(2013)\citenamefont
  {Ferguson} \emph {et~al.}}]{ferguson2012hivtrap}%
  \BibitemOpen
  \bibfield  {author} {\bibinfo {author} {\bibfnamefont {A.~L.}\ \bibnamefont
  {Ferguson}} \emph {et~al.},\ }\bibfield  {title} {\enquote {\bibinfo {title}
  {Translating {HIV} sequences into quantitative fitness landscapes predicts
  viral vulnerabilities for rational immunogen design},}\ }\href@noop {}
  {\bibfield  {journal} {\bibinfo  {journal} {\emph{Immunity}}\ }\textbf
  {\bibinfo {volume} {38}},\ \bibinfo {pages} {606--617} (\bibinfo {year}
  {2013})}\BibitemShut {NoStop}%
\bibitem [{\citenamefont {Jaynes}(1957)}]{jaynes1957infotheory}%
  \BibitemOpen
  \bibfield  {author} {\bibinfo {author} {\bibfnamefont {E.~T.}\ \bibnamefont
  {Jaynes}},\ }\bibfield  {title} {\enquote {\bibinfo {title} {Information
  theory and statistical mechanics},}\ }\href@noop {} {\bibfield  {journal}
  {\bibinfo  {journal} {\emph{Physical Review}}\ }\textbf {\bibinfo {volume}
  {106}},\ \bibinfo {pages} {620--630} (\bibinfo {year} {1957})}\BibitemShut
  {NoStop}%
\bibitem [{\citenamefont {Tkacik}\ \emph {et~al.}(2009)\citenamefont {Tkacik},
  \citenamefont {Schneidman}, \citenamefont {II},\ and\ \citenamefont
  {Bialek}}]{tkacik2009spin}%
  \BibitemOpen
  \bibfield  {author} {\bibinfo {author} {\bibfnamefont {G.}~\bibnamefont
  {Tkacik}}, \bibinfo {author} {\bibfnamefont {E.}~\bibnamefont {Schneidman}},
  \bibinfo {author} {\bibfnamefont {M.~J.~Berry}\ \bibnamefont {II}}, \ and\
  \bibinfo {author} {\bibfnamefont {W.}~\bibnamefont {Bialek}},\ }\bibfield
  {title} {\enquote {\bibinfo {title} {Spin glass models for a network of real
  neurons},}\ }\href@noop {} {\bibfield  {journal} {\bibinfo  {journal}
  {\emph{arXiv preprint arXiv:0912.5409}}\ } (\bibinfo {year}
  {2009})}\BibitemShut {NoStop}%
\bibitem [{hiv()}]{hivdb}%
  \BibitemOpen
  \href@noop {} {\enquote {\bibinfo {title} {{Los} {Alamos} {HIV} {Sequence}
  {Database}},}\ }\bibinfo {howpublished}
  {\url{http://www.hiv.lanl.gov/}}\BibitemShut {NoStop}%
\bibitem [{\citenamefont {Binder}\ and\ \citenamefont
  {Young}(1986)}]{binder1986spin}%
  \BibitemOpen
  \bibfield  {author} {\bibinfo {author} {\bibfnamefont {K.}~\bibnamefont
  {Binder}}\ and\ \bibinfo {author} {\bibfnamefont {A.~P.}\ \bibnamefont
  {Young}},\ }\bibfield  {title} {\enquote {\bibinfo {title} {Spin glasses:
  Experimental facts, theoretical concepts, and open questions},}\ }\href@noop
  {} {\bibfield  {journal} {\bibinfo  {journal} {\emph{Reviews of Modern
  physics}}\ }\textbf {\bibinfo {volume} {58}},\ \bibinfo {pages} {801}
  (\bibinfo {year} {1986})}\BibitemShut {NoStop}%
\bibitem [{\citenamefont {Sella}\ and\ \citenamefont
  {Hirsh}()}]{sella2005stat}%
  \BibitemOpen
  \bibfield  {author} {\bibinfo {author} {\bibfnamefont {G.}~\bibnamefont
  {Sella}}\ and\ \bibinfo {author} {\bibfnamefont {A.~E.}\ \bibnamefont
  {Hirsh}},\ }\bibfield  {title} {\enquote {\bibinfo {title} {The application
  of statistical physics to evolutionary biology},}\ }\href@noop {} {\bibfield
  {journal} {\bibinfo  {journal} {\emph{Proc. Natl. Acad. Sci.}}\ }\textbf
  {\bibinfo {volume} {102}},\ \bibinfo {pages} {9541--9546}}\BibitemShut
  {NoStop}%
\bibitem [{\citenamefont {Brumme}\ \emph {et~al.}(2009)\citenamefont {Brumme}
  \emph {et~al.}}]{brumme2009pathway}%
  \BibitemOpen
  \bibfield  {author} {\bibinfo {author} {\bibfnamefont {Z.~L.}\ \bibnamefont
  {Brumme}} \emph {et~al.},\ }\bibfield  {title} {\enquote {\bibinfo {title}
  {{HLA}-associated immune escape pathways in {HIV}-1 subtype {B} {Gag}, {Pol}
  and {Nef} proteins},}\ }\href@noop {} {\bibfield  {journal} {\bibinfo
  {journal} {\emph{PLoS one}}\ }\textbf {\bibinfo {volume} {4}},\ \bibinfo
  {pages} {e6687} (\bibinfo {year} {2009})}\BibitemShut {NoStop}%
\bibitem [{\citenamefont {Turner}\ and\ \citenamefont
  {Summers.}(1999)}]{turner1999struc}%
  \BibitemOpen
  \bibfield  {author} {\bibinfo {author} {\bibfnamefont {B.~G.}\ \bibnamefont
  {Turner}}\ and\ \bibinfo {author} {\bibfnamefont {M.~F.}\ \bibnamefont
  {Summers.}},\ }\bibfield  {title} {\enquote {\bibinfo {title} {Structural
  biology of {HIV}},}\ }\href@noop {} {\bibfield  {journal} {\bibinfo
  {journal} {\emph{Journal of Molecular Biology}}\ }\textbf {\bibinfo {volume}
  {285}},\ \bibinfo {pages} {1--32} (\bibinfo {year} {1999})}\BibitemShut
  {NoStop}%
\bibitem [{\citenamefont {Rouzine}\ and\ \citenamefont
  {Coffin}(1999)}]{rouzine1999linkage}%
  \BibitemOpen
  \bibfield  {author} {\bibinfo {author} {\bibfnamefont {I.~M.}\ \bibnamefont
  {Rouzine}}\ and\ \bibinfo {author} {\bibfnamefont {J.~M.}\ \bibnamefont
  {Coffin}},\ }\bibfield  {title} {\enquote {\bibinfo {title} {Linkage
  disequilibrium test implies a large effective population number for {HIV} in
  vivo},}\ }\href@noop {} {\bibfield  {journal} {\bibinfo  {journal}
  {\emph{Proc. Natl. Acad. Sci.}}\ }\textbf {\bibinfo {volume} {96}},\ \bibinfo
  {pages} {10758--10763} (\bibinfo {year} {1999})}\BibitemShut {NoStop}%
\bibitem [{\citenamefont {Kouyos}\ \emph {et~al.}(2006)\citenamefont {Kouyos},
  \citenamefont {Althaus},\ and\ \citenamefont
  {Bonhoeffer}}]{kouyos2006stochastic}%
  \BibitemOpen
  \bibfield  {author} {\bibinfo {author} {\bibfnamefont {R.~D.}\ \bibnamefont
  {Kouyos}}, \bibinfo {author} {\bibfnamefont {C.~L.}\ \bibnamefont {Althaus}},
  \ and\ \bibinfo {author} {\bibfnamefont {S.}~\bibnamefont {Bonhoeffer}},\
  }\bibfield  {title} {\enquote {\bibinfo {title} {Stochastic or deterministic:
  what is the effective population size of {HIV}-1?}}\ }\href@noop {}
  {\bibfield  {journal} {\bibinfo  {journal} {\emph{Trends in microbiology}}\
  }\textbf {\bibinfo {volume} {14}},\ \bibinfo {pages} {507--511} (\bibinfo
  {year} {2006})}\BibitemShut {NoStop}%
\bibitem [{\citenamefont {Eigen}(1971)}]{eigen1971selforganization}%
  \BibitemOpen
  \bibfield  {author} {\bibinfo {author} {\bibfnamefont {M.}~\bibnamefont
  {Eigen}},\ }\bibfield  {title} {\enquote {\bibinfo {title} {Selforganization
  of matter and the evolution of biological macromolecules},}\ }\href@noop {}
  {\bibfield  {journal} {\bibinfo  {journal} {\emph{Naturwissenschaften}}\
  }\textbf {\bibinfo {volume} {58}},\ \bibinfo {pages} {465--523} (\bibinfo
  {year} {1971})}\BibitemShut {NoStop}%
\bibitem [{\citenamefont {Bonhoeffer}\ \emph {et~al.}(2004)\citenamefont
  {Bonhoeffer}, \citenamefont {Chappey}, \citenamefont {Parkin}, \citenamefont
  {Whitcomb},\ and\ \citenamefont {Petropoulos}}]{bonhoeffer2004evidence}%
  \BibitemOpen
  \bibfield  {author} {\bibinfo {author} {\bibfnamefont {S.}~\bibnamefont
  {Bonhoeffer}}, \bibinfo {author} {\bibfnamefont {C.}~\bibnamefont {Chappey}},
  \bibinfo {author} {\bibfnamefont {N.~T.}\ \bibnamefont {Parkin}}, \bibinfo
  {author} {\bibfnamefont {J.~M.}\ \bibnamefont {Whitcomb}}, \ and\ \bibinfo
  {author} {\bibfnamefont {C.~J.}\ \bibnamefont {Petropoulos}},\ }\bibfield
  {title} {\enquote {\bibinfo {title} {Evidence for positive epistasis in
  {HIV}-1},}\ }\href@noop {} {\bibfield  {journal} {\bibinfo  {journal}
  {\emph{Science}}\ }\textbf {\bibinfo {volume} {306}},\ \bibinfo {pages}
  {1547--1550} (\bibinfo {year} {2004})}\BibitemShut {NoStop}%
\bibitem [{\citenamefont {Leuth\"{a}usser}(1987)}]{leuthausser1987eigen}%
  \BibitemOpen
  \bibfield  {author} {\bibinfo {author} {\bibfnamefont {I.}~\bibnamefont
  {Leuth\"{a}usser}},\ }\bibfield  {title} {\enquote {\bibinfo {title}
  {Statistical mechanics of {Eigen}'s evolution model},}\ }\href@noop {}
  {\bibfield  {journal} {\bibinfo  {journal} {\emph{J. Stat. Phys.}}\ }\textbf
  {\bibinfo {volume} {48}},\ \bibinfo {pages} {343--360} (\bibinfo {year}
  {1987})}\BibitemShut {NoStop}%
\bibitem [{\citenamefont {Feynman}\ and\ \citenamefont
  {Hibbs}(1965)}]{feynman1965quantum}%
  \BibitemOpen
  \bibfield  {author} {\bibinfo {author} {\bibfnamefont {R.~P.}\ \bibnamefont
  {Feynman}}\ and\ \bibinfo {author} {\bibfnamefont {A.~R.}\ \bibnamefont
  {Hibbs}},\ }\href@noop {} {\emph {\bibinfo {title} {Quantum Mechanics and
  Path integrals}}}\ (\bibinfo  {publisher} {MacGraw Hill},\ \bibinfo {year}
  {1965})\BibitemShut {NoStop}%
\bibitem [{\citenamefont {Dixit}\ \emph {et~al.}(2012)\citenamefont {Dixit},
  \citenamefont {Srivastava},\ and\ \citenamefont {Vishnoi}}]{dixit2012finite}%
  \BibitemOpen
  \bibfield  {author} {\bibinfo {author} {\bibfnamefont {N.M.}\ \bibnamefont
  {Dixit}}, \bibinfo {author} {\bibfnamefont {P.}~\bibnamefont {Srivastava}}, \
  and\ \bibinfo {author} {\bibfnamefont {N.~K.}\ \bibnamefont {Vishnoi}},\
  }\bibfield  {title} {\enquote {\bibinfo {title} {A finite population model of
  molecular evolution: {T}heory and computation},}\ }\href@noop {} {\bibfield
  {journal} {\bibinfo  {journal} {\emph{Journal of Computational Biology}}\
  }\textbf {\bibinfo {volume} {19}},\ \bibinfo {pages} {1176--1202} (\bibinfo
  {year} {2012})}\BibitemShut {NoStop}%
\bibitem [{dix()}]{dixit2012note}%
  \BibitemOpen
  \href@noop {} {}\bibinfo {note} {In \cite{dixit2012finite}, the authors
  simulate a finite population model called the ``RSM'' model, that is
  structurally similar to the model of intra host evolution we have considered
  in this paper.}\BibitemShut {Stop}%
\bibitem [{\citenamefont {Corder}\ and\ \citenamefont
  {Foreman}(2009)}]{corder2009nonparametric}%
  \BibitemOpen
  \bibfield  {author} {\bibinfo {author} {\bibfnamefont {G.W.}\ \bibnamefont
  {Corder}}\ and\ \bibinfo {author} {\bibfnamefont {D.I.}\ \bibnamefont
  {Foreman}},\ }\href@noop {} {\emph {\bibinfo {title} {Nonparametric
  Statistics for Non-statisticians: A Step-by-step Approach}}}\ (\bibinfo
  {publisher} {Wiley},\ \bibinfo {year} {2009})\BibitemShut {NoStop}%
\bibitem [{\citenamefont {Henn}\ \emph {et~al.}(2012)\citenamefont {Henn} \emph
  {et~al.}}]{henn2012whole}%
  \BibitemOpen
  \bibfield  {author} {\bibinfo {author} {\bibfnamefont {M.R.}\ \bibnamefont
  {Henn}} \emph {et~al.},\ }\bibfield  {title} {\enquote {\bibinfo {title}
  {Whole genome deep sequencing of {HIV-1} reveals the impact of early minor
  variants upon immune recognition during acute infection},}\ }\href@noop {}
  {\bibfield  {journal} {\bibinfo  {journal} {\emph{PLoS pathogens}}\ }\textbf
  {\bibinfo {volume} {8}},\ \bibinfo {pages} {e1002529} (\bibinfo {year}
  {2012})}\BibitemShut {NoStop}%
\bibitem [{\citenamefont {Amitrano}\ \emph {et~al.}(1989)\citenamefont
  {Amitrano}, \citenamefont {Peliti},\ and\ \citenamefont
  {Saber}}]{amitrano1989evol}%
  \BibitemOpen
  \bibfield  {author} {\bibinfo {author} {\bibfnamefont {C.}~\bibnamefont
  {Amitrano}}, \bibinfo {author} {\bibfnamefont {L.}~\bibnamefont {Peliti}}, \
  and\ \bibinfo {author} {\bibfnamefont {M.}~\bibnamefont {Saber}},\ }\bibfield
   {title} {\enquote {\bibinfo {title} {Population dynamics in a spin-glass
  model of chemical evolution},}\ }\href@noop {} {\bibfield  {journal}
  {\bibinfo  {journal} {\emph{Journal of Molecular Evolution}}\ }\textbf
  {\bibinfo {volume} {29}},\ \bibinfo {pages} {513--525} (\bibinfo {year}
  {1989})}\BibitemShut {NoStop}%
\bibitem [{\citenamefont {Metropolis}\ \emph {et~al.}(1953)\citenamefont
  {Metropolis}, \citenamefont {Rosenbluth}, \citenamefont {Rosenbluth},
  \citenamefont {Teller},\ and\ \citenamefont
  {Teller}}]{metropolis1953equation}%
  \BibitemOpen
  \bibfield  {author} {\bibinfo {author} {\bibfnamefont {N.}~\bibnamefont
  {Metropolis}}, \bibinfo {author} {\bibfnamefont {A.~W.}\ \bibnamefont
  {Rosenbluth}}, \bibinfo {author} {\bibfnamefont {M.~N.}\ \bibnamefont
  {Rosenbluth}}, \bibinfo {author} {\bibfnamefont {A.H.}\ \bibnamefont
  {Teller}}, \ and\ \bibinfo {author} {\bibfnamefont {E.}~\bibnamefont
  {Teller}},\ }\bibfield  {title} {\enquote {\bibinfo {title} {Equation of
  state calculations by fast computing machines},}\ }\href@noop {} {\bibfield
  {journal} {\bibinfo  {journal} {\emph{The Journal of Chemical Physics}}\
  }\textbf {\bibinfo {volume} {21}},\ \bibinfo {pages} {1087} (\bibinfo {year}
  {1953})}\BibitemShut {NoStop}%
\bibitem [{\citenamefont {Friedman}\ \emph {et~al.}(2001)\citenamefont
  {Friedman}, \citenamefont {Hastie},\ and\ \citenamefont
  {Tibshirani}}]{friedman2001elements}%
  \BibitemOpen
  \bibfield  {author} {\bibinfo {author} {\bibfnamefont {J.}~\bibnamefont
  {Friedman}}, \bibinfo {author} {\bibfnamefont {T.}~\bibnamefont {Hastie}}, \
  and\ \bibinfo {author} {\bibfnamefont {R.}~\bibnamefont {Tibshirani}},\
  }\href@noop {} {\emph {\bibinfo {title} {The Elements of Statistical
  Learning}}}\ (\bibinfo  {publisher} {Springer Series in Statistics},\
  \bibinfo {year} {2001})\BibitemShut {NoStop}%
\end{thebibliography}%

\end{document}